\documentstyle[prl,aps]{revtex}
\input epsf.sty

\begin{document}
\draft
\twocolumn[\columnwidth\textwidth\csname@twocolumnfalse\endcsname

\title{Odd-Even Staggering of  Nuclear Masses: Pairing or Shape Effect?}

\author{
W. Satu{\l}a,$^{1-3}$
J. Dobaczewski,$^{1-3}$ and
W. Nazarewicz$^{2-4}$
}

\address{$^1$Joint Institute for Heavy Ion Research, Oak Ridge,
             Tennessee 37831}
\address{$^2$Department of Physics, University of Tennessee, Knoxville,
             Tennessee 37996}
\address{$^3$Institute of Theoretical Physics, University of Warsaw,
             ul. Ho\.za 69, PL-00-681 Warsaw, Poland}
\address{$^4$Physics Division, Oak Ridge National Laboratory,
                Oak Ridge, Tennessee 37831}

\date{\today}

\maketitle

\begin{abstract}
The odd-even  staggering
of nuclear masses
 was recognized in the early
days of nuclear physics. Recently, a similar effect was
discovered in other finite fermion systems, such as
ultrasmall metallic grains and metal clusters.
It is  believed that the staggering
in nuclei and grains
is primarily due to pairing  correlations (superconductivity), while
in clusters it is caused by the Jahn-Teller effect.
We find that, for light and medium-mass nuclei,
 the  staggering has two components.
The first one originates from
pairing while  the second, comparable in magnitude,
has its roots in   the deformed
mean field.
\end{abstract}

\pacs{PACS number(s): 21.10.Dr, 21.10.Pc, 21.60.Jz, 71.15.Mb}
\addvspace{5mm}]

\narrowtext

The odd-even staggering (OES) of binding energies
has  been observed in several
finite many-fermion systems such as
nuclei  \cite{[Hei32]},
ultrasmall superconducting
grains \cite{[Bla96]}, and  metal clusters \cite{[deH93]}.
It manifests itself in the fact that the binding
energy  of
a system with an odd particle number  is lower than
the arithmetic mean of the energies of the two neighboring
even-particle-number systems.

In atomic nuclei, the OES
is usually attributed
to the existence of nucleonic pairing correlations
\cite{[Boh58],[Boh69]}. A similar scenario
has been proposed for metallic grains (see
Refs. \cite{[Ros98],[Bal98]} and references quoted therein). In
both cases, the finite-size effects are important, and
the Cooper pairing is well described in terms
of the parity-number-conserving quasiparticle approach.
Although the motion of electrons
in metals is very different from that of nucleons in nuclei,
the mechanism behind electronic and nucleonic
superconductivity
(presence of  attractive residual interaction which gives rise
to a correlated many-fermion system) is indeed
very similar \cite{[Boh75],[Lip98]}.

So far,  no evidence has been found for
superconductivity
in alkali metal clusters \cite{[Kuz96]}, and
the OES  of binding energies in such systems is
attributed to  a very
different mechanism. Namely, it is believed to have
 its origin in
the Jahn-Teller effect which, by breaking the spherical symmetry
of the mean field, gives rise to
deformed  single-particle  orbitals
\cite{[Cle85],[Man94]}.
Recently,
H\"akkinen {\em et al.}  \cite{[Hak97]},
using the density-functional theory,  argued that
light alkali-metal clusters and light $N$=$Z$ nuclei
have a similar pattern of OES, irrespective of differences in the
interactions between the fermions. Hence,
they concluded that
the OES in small nuclei appears to be a
mere deformation effect rather
than a consequence of pairing.

The main objective of this study is to analyze
 the phenomenon of OES in nuclei
from the microscopic perspective.
Guided by self-consistent calculations,
we make  an attempt
to determine and separate the pairing
and mean-field contributions to the experimental OES.

In the independent  quasiparticle (BCS) picture
\cite{[Bar57]}, the gap parameter,
$\Delta$, can be related to the binding energies of  three
 adjacent systems.
 Assuming that the binding energies of even systems and
those of odd systems are locally
smooth
functions of the particle number $N$,
the quantity
\begin{equation}\label{d3}
\Delta^{(3)}(N) \equiv
{\pi_N \over 2}  [B(N-1) + B(N+1) -2B(N)]
\end{equation}
is often interpreted as a measure of the empirical pairing gap.
In Eq.~(\ref{d3}),
$\pi_N = (-1)^N$
is the  number parity and $B(N)$ is the (negative) binding energy
of a system with $N$ particles.
Indeed, assuming that the ground state of the
odd-$N$ system is a pure one-quasiparticle state,
one has
$\Delta^{(3)}(N)$$\approx$$E_k$$\approx$$\Delta$, where
$E_k$=$\sqrt{(e_k-\lambda)^2+\Delta^2}$ is the
lowest BCS quasiparticle energy,
$e_k$ is the energy of the
single-particle  orbital occupied by the odd
nucleon, and
$\lambda$ (=$dB/dN$) is the Fermi energy.

Another  commonly used binding-energy relation
for  $\Delta$ is the
four-point expression \cite{[Nil61]},
\begin{eqnarray}
\Delta^{(4)}(N) & \equiv &
{\pi_N\over 4}  \left[3B(N-1)-3 B(N)-B(N-2)\right. \nonumber \\
&+&\left.B(N+1)\right]
=\frac{1}{2} [\Delta^{(3)}(N) + \Delta^{(3)}(N-1)]  \label{d34}
,
\end{eqnarray}
which averages the $\Delta^{(3)}$ values in adjacent even and odd
systems. 
In nuclei, because there
are two kinds of particles, neutrons and protons,
$\Delta$ is calculated
along the isotopic or isotonic chains \cite{[Boh69]}.
The resulting pairing
gaps are denoted as $\Delta_\nu$ and $\Delta_\pi$,
respectively.
In what follows we argue  that different physical effects
determine the behavior of 
$\Delta^{(3)}(N)$ for odd and even particle numbers. Hence,
the interpretation of the average (\ref{d34}) 
in terms of the pairing gap can be misleading.

To investigate the Jahn-Teller component of the OES, we performed the
Hartree-Fock (HF) calculations without pairing for
nuclei with $Z$=9$\div$28 and
$N$$-Z$=$-2$$\div$6.
The HFODD code (v1.75) \cite{[Dob97]} and two
different Skyrme  parametrizations, SIII \cite{[Bei75]}
 and SLy4
 \cite{[Cha97]}, were employed.
For each Skyrme force,
two  sets of calculations were carried out:
either requiring time-reversal symmetry to be conserved or not.
(The time-odd components
appear in the mean fields of odd and odd-odd systems.)
 Since, according to
our calculations, the time-odd terms
do not affect {\em qualitatively} the OES,
the results presented in this work
were obtained by  neglecting  these terms.

\begin{figure}
\begin{center}
\leavevmode
\epsfysize=10.6cm
\epsfbox{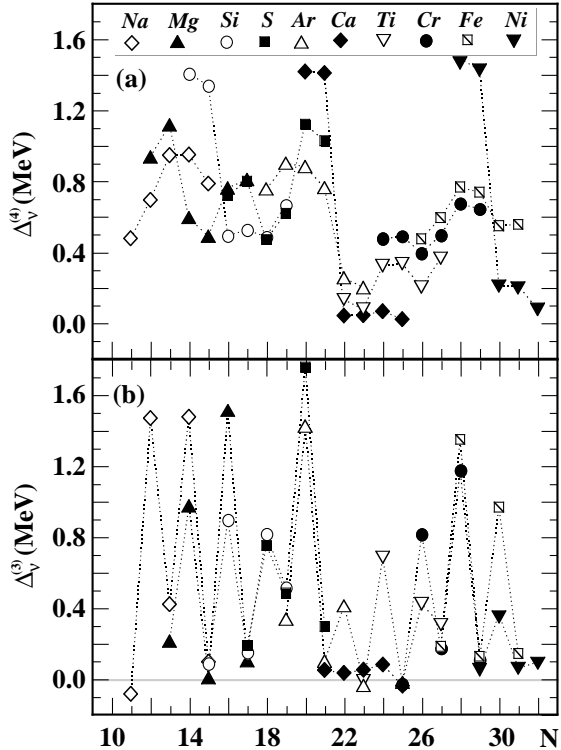}
\end{center}
\caption{$\Delta^{(4)}_\nu$ (a)
and $\Delta^{(3)}_\nu$ (b)  calculated
in the SIII HF model
without pairing as a function of neutron number.
}
\label{pd4}
\end{figure}
{}Figure \ref{pd4}a displays the theoretical values of
$\Delta^{(4)}_\nu$
 obtained
from  binding energies in the
SIII model. The results obtained with the
SLy4 parametrization  are strikingly  similar,
in spite of the fact that the total binding
energies predicted by these two forces show large differences,
approaching 4\,MeV in some cases.
This result suggests that,
unlike the total binding energy,
$\Delta^{(4)}$  weakly depends on
the effective interaction.

Results shown in Fig.~\ref{pd4}a
demonstrate that the self-consistent mean-field
theory  {\em without} pairing  does
indeed  predict
the OES according to the criterion (\ref{d34}).
The effect is sizable: theoretical values of $\Delta^{(4)}_\nu$
reach 30\% to 50\% of the empirical OES and they, on average,
gradually decrease with mass.
A rather complicated pattern of $\Delta^{(4)}_\nu$ can be easily
explained by looking at values of $\Delta^{(3)}_\nu$
presented in Fig.~\ref{pd4}b. Values of $\Delta^{(3)}_\nu$ are
large for $\pi_N$=+1  and very small for $\pi_N$=--1; hence
the averages $\Delta^{(4)}_\nu$ (\ref{d34}) simply reflect the simple pattern of
$\Delta^{(3)}_\nu$.
(The behavior of $\Delta^{(3)}_\pi$
 shows a very similar pattern. Also
much the same results were obtained with  the SLy4 force.)

We are now in a position to trace the pattern of
$\Delta^{(3)}_\nu$, shown in Fig.~\ref{pd4}b, 
back to properties
of the deformed mean field.
Indeed,
Eq.~(\ref{d3}) represents the second-order difference
 with respect to
$N$, i.e.,
\begin{equation}\label{der}
2\pi_N\Delta^{(3)}(N) \approx
\frac{\partial^2 B}{\partial N^2}=
\frac{\partial \lambda}{\partial N}
= {1\over g(\lambda)},
\end{equation}
where $g(e)$$\equiv$$dN/de$ is the single-particle level density.
Consequently, in the absence of  OES due to pairing,
$\Delta^{(3)}$ represents the variation of the Fermi energy with
particle number. In the case of a degenerate shell, the Fermi energy
does not change with $N$ ($\lambda$ lies on the last occupied level)
 and $\Delta^{(3)}$=0. The change in $\lambda$ takes place when
the valence shell $e_n$ ($N$=$2n$)
is completely filled and the higher shell,
$e_{n+1}$, needs to be occupied. In  this case, corresponding
to $\pi_N$=+1,
$d\lambda/dN$$\approx$$e_{n+1}$--$e_n$, and
$\Delta^{(3)}$$\approx$$(e_{n}$--$e_{n+1})/2$.
That is, in the absence of pairing correlations, $\Delta^{(3)}$
becomes a measure of a   gap in the single-particle spectrum.
This single-particle
mechanism behind the OES was early noticed in Ref.~\cite{[Boh69]}
and subsequently employed  in Refs.~\cite{[Cle85],[Man94]}
to explain
the OES in metal clusters. The alternating behavior of  $\Delta^{(3)}$
in Fig.~\ref{pd4}b comes from the twofold Kramers degeneracy
of the deformed single-particle energy levels. Indeed,
the spherical symmetry of the mean-field potential, which
gives rise to a ($2j$+1)-fold degeneracy of
single-particle levels,  is preserved only
for doubly magic nuclei.
For open-shell systems, spherical
symmetry is spontaneously broken by
the Jahn-Teller mechanism, and
the ground state is characterized by
the deformed mean field, cf.~Refs. \cite{[Rei84],[Naz94a]}.

Results of the self-consistent calculations
for nuclei and metal clusters can
be understood in terms of the macroscopic-microscopic
 shell-correction method. In this approach,
which can be viewed as an approximation to the
HF treatment
\cite{[Str74]},
the total binding energy can be written as
$B=E_{\rm sp} - \tilde{E}_{\rm sp} +E_{\rm macro}$,
where
\begin{equation}\label{sps}
E_{\rm sp} = {\textstyle{\sum_{k=1}^{A}e_k}}
\end{equation}
is the shell-model energy (sum of single-particle energies of occupied states),
$\tilde{E}_{\rm sp}$ is the Strutinsky-averaged shell-model energy,
and $E_{\rm macro}$ stands for the macroscopic 
liquid-drop energy.

The liquid-drop contributions to the second difference
(\ref{der}) differ for nuclei and clusters. For nuclei,
the main effect comes from the symmetry energy term
\cite{[Nem62]}.
Assuming
the symmetry energy coefficient
$a_I$=38\,MeV (the value
appropriate for light nuclei \cite{[Sat98]})
the nuclear liquid-drop curvature contribution to (\ref{der}) is
38$/A$\,MeV, while the contribution of the surface-energy
term is much smaller.
On the contrary, in alkali-metal clusters, the {\em
leading}
contribution from the
liquid-drop term to (\ref{der}) comes from
the surface energy and is negligible.
Indeed, taking  the typical value
of the surface-tension
coefficient for the bulk Na,
$\sigma$$\approx$0.01\,eV/{\AA}$^2$, one obtains
a very small correction
 $\partial^2E_{\rm macro}/\partial N^2\approx-0.15/N^{4/3}$\,eV.

In order to evaluate the curvature contribution
from  $\tilde{E}_{\rm sp}$, one needs
 to estimate the average single-particle level
density at the Fermi energy, $g(\lambda)$ [see Eq.~(\ref{der})].
In the nuclear case $g(\lambda)=3a/\pi^2$,
where $a$
is the level density parameter \cite{[Boh69]}.
Experimentally, $a$$\approx$$A$/8\,MeV for light nuclei,
and this agrees well with the estimate
based on  realistic
potentials \cite{[Shl92]}.
Since in our HF calculations the effective mass is less than
 one,
the level density parameter should
be additionally multiplied by the effective-mass
factor: $m^*/m$=0.76  and 0.7 for SIII and SLy4, respectively.
Consequently, the corresponding curvature contribution becomes
$-1/g(\lambda)$$\approx$$-36/A$\,MeV, and
nearly cancels out the liquid-drop contribution.
One  can, therefore, conclude
that the leading contribution to the HF values of
$\Delta^{(3)}$ shown in Fig.~\ref{pd4}b
comes from the single-particle
sum (\ref{sps}),
\begin{equation}\label{eeff}
\delta e\equiv \Delta^{(3)}_{\rm sp}(N) \approx
 {1\over 4} (1+\pi_N)(e_{n+1}-e_n),
\end{equation}
provided that one neglects  small
shifts in  the  single-particle energies
due to variations of the mean field (e.g., deformation changes)
when increasing the particle number from $N$ to
$N$+1. (The effect of the mean-field variations is seen
in Fig.~\ref{pd4}b for particle numbers 13, 19, and 27, i.e.,
around magic gaps where the shape transitions take place.)
Quantity (\ref{eeff}) represents, therefore, the {\em effective}
single-particle spacing between nucleonic energy levels.

The above conclusion
is not true for alkali-metal
clusters.
The level density for alkali-metal clusters can be estimated
following Ref.~\cite{[Boh69]}:
$g(\lambda)=0.154 m_er_0^2N/\hbar^2$. By taking
$r_0$=2.17\,{\AA} (the value
corresponding to the density of bulk Na at 500$^\circ$K), one
obtains $1/g({\lambda})$$\approx$1.4$/N$\,eV.
(This result is fairly close to the Fermi-gas estimate
 of $1/g({\lambda})$$\approx$2.15$/N$\,eV,
assuming $\lambda$=3.23\,eV.)
Contrary to the nuclear case, since the liquid-drop component of
the second difference is very small, the
smoothed-energy  term strongly
contributes to $\Delta^{(3)}$. Moreover, the typical
single-particle splitting due to cluster deformation is of the
order of 0.2\,eV \cite{[Fra96]}, i.e., it is very similar
to the value of $1/g({\lambda})$.
For example, in Fig.~4 of Ref.~\cite{[Hak97]},
the calculated OES parameter $2\pi_N\Delta^{(3)}$
oscillates around {\em zero} - in contrast to the nuclear
case presented in Fig.~\ref{pd4}b. This is because
the  smoothed-energy contribution to the OES, $-1/g({\lambda})$,
introduces  a negative shift of the same order as
term (\ref{eeff}).

\begin{figure}
\begin{center}
\leavevmode
\epsfysize=5.2cm
\epsfbox{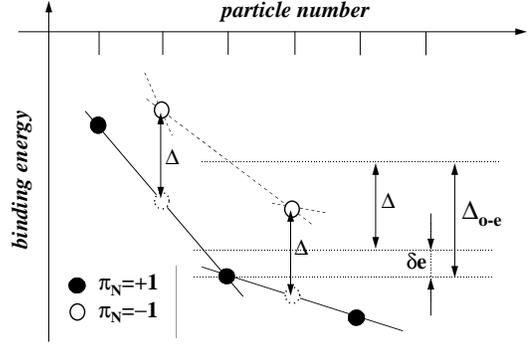}
\end{center}
\caption{Schematic illustration
of various contributions to the OES. The odd-even energy
difference, $\Delta_{\rm o-e}$, is decomposed into the
 pairing
part, $\Delta$, and the mean-field part, $\delta e$.
}\label{delta3}
\end{figure}

Since in nuclei the mean-field contribution $\delta e$
is very small for $\pi_N$=$-1$,
both pairing and mean-field  components of OES
can be  extracted from  binding energies by using the
three-point filter $\Delta^{(3)}$.
As illustrated schematically in Fig.~\ref{delta3},
values of $\Delta^{(3)}$ calculated at odd values of $N$ 
can be associated with
the pairing effect \cite{sen},
\begin{equation}\label{pairing}
\Delta_\nu(N)\equiv\Delta_\nu^{(3)}(N=2n+1),
\end{equation}
while the differences of $\Delta^{(3)}$ at adjacent even and odd
values of $N$ give information about the single-particle spectra,
\begin{equation}\label{nilsson}
e_{n+1} - e_{n} =2\left[\Delta_\nu^{(3)}(N=2n) - \Delta_\nu^{(3)}(N=2n+1)
\right].
\end{equation}

\begin{figure}
\begin{center}
\leavevmode
\epsfysize=6.0cm
\epsfbox{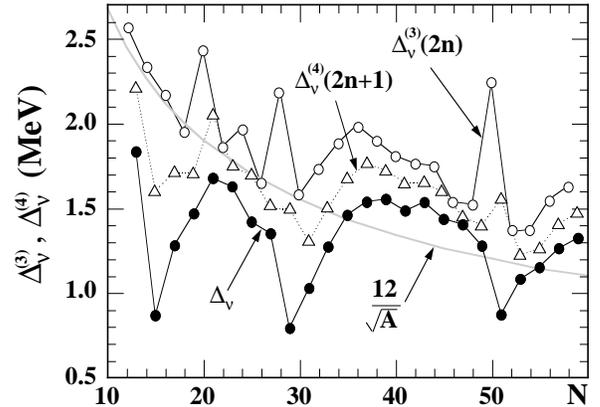}
\end{center}
\caption{Experimental values of
$\Delta_\nu(N)=\Delta_\nu^{(3)}(N)$ for $N$-odd
(filled circles),
$\Delta_\nu^{(3)}$ for $N$-even (open circles),
and $\Delta_\nu^{(4)}$ for $N$-odd (open triangles).
The thick gray line
indicates the average trend,
 $\tilde\Delta$=12/$\protect\sqrt{A}$.
Each  point represents  the arithmetic mean over several
even-$Z$ isotones.
   }
\label{expgap}
\end{figure}
The neutron pairing gaps (\ref{pairing}) extracted from
the experimental binding energies
are shown in Fig.~\ref{expgap}.
The expected quenching of neutron pairing
 at magic (or semi-magic) particle numbers $N$=14, 28, and 50
is clearly seen. (Interestingly, the
 minimum at $N$=20 is absent.)

The experimental values of
$\Delta_\nu^{(3)}$  at even neutron numbers 
are systematically shifted with respect to $\Delta_\nu$. Since the
differences (\ref{nilsson})
  reflect the mean-field contributions
to the OES, they sharply peak at
magic numbers.
In the experimental values of
$\Delta_\nu^{(4)}$  the magic gaps are almost invisible.
This is so because the effects of a large
single-particle gap and 
quenched pairing cancel out in the averages
(\ref{d34}).
Since the
commonly used smooth dependence of average pairing gap on mass number,
$\tilde\Delta=12/\sqrt{A}$, was fitted to $\Delta^{(4)}$,
and  not
to $\Delta$
\cite{[Zel67]}, the values of
$\tilde\Delta$ are overestimated, especially for light systems.
As seen in Fig.~\ref{expgap}, the values of  $\Delta_\nu$
in the middle of the $sd$ and $pf$ shells
are fairly similar, and are not consistent with the
${A}^{-1/2}$ trend.

\begin{figure}
\begin{center}
\leavevmode
\epsfysize=4.2cm
\epsfbox{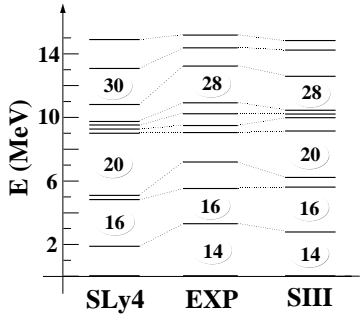}
\end{center}
\caption{Experimental and calculated
 effective single-particle spectrum extracted by means
of relation (\protect\ref{nilsson})
for  $N$$-Z$=4.
         }
\label{level}
\end{figure}
Finally, we have extracted from the data
 the mean-field contributions
to the OES
 according to Eq.~(\ref{nilsson}). As an example,
in Fig.~\ref{level} they are compared with the HF results
 for the  $N$$-Z$=4 chains.
In spite of the fact that no pairing
correlations have been considered in calculations,
a good agreement between the experimental and
theoretical levels is obtained.
The fact that theoretical and experimental energy scales agree
is a consequence of the fact that
both SIII and SLy4
reproduce fairly well the symmetry energy.
The  level bunching predicted 
 between the $N$=20 and $N$=28 gaps reflects
small calculated equilibrium deformations  in
these $1f_{7/2}$-shell nuclei, which nicely agrees with a similar
grouping of levels seen in  experiment.

In summary,
our analysis does not confirm the recent suggestion \cite{[Hak97]}
that the OES in light nuclei is a mere reflection of the deformed mean
field.
We have demonstrated that
the OES  in light atomic nuclei
is strongly affected by {\em both} nucleonic
pairing {\em and} the deformed mean-field.
A method has been proposed to extract the pairing contribution to the
OES from experimental data. The experimental pairing gaps
show a weaker $A$-dependence than that
obtained previously.
 Since the fourth-order mass difference $\Delta^{(4)}$
is strongly affected by the mean-field contribution, it
is not a good measure of pairing correlations, at least in  light nuclei.

In our discussion we have not discussed the
singular behavior of binding energies of the
$N$=$Z$ nuclei, known as
the  Wigner energy, which dramatically influences
the binding-energy relations (\ref{d3}) and (\ref{d34})
\cite{[Jen84],[Sat97],[Sat98]}. Consequently,
the OES near the $N$=$Z$
line  has an additional {\em third} component 
originating from the neutron-proton correlations.  In particular,
the OES
parameter discussed in Ref.~\cite{[Hak97]}, which is based
on binding energy differences of even-even and odd-odd
$N$=$Z$ nuclei
(hence it {\em does not represent} any pairing gap),
is strongly perturbed by this third component
 \cite{[Sat97]}.


Useful discussions with S. Frauendorf
are gratefully acknowledged.
This research was supported in part by
the U.S. Department of Energy
under Contract Nos. DE-FG02-96ER40963 (University of Tennessee),
DE-FG05-87ER40361 (Joint Institute for Heavy Ion Research),
DE-AC05-96OR22464 with Lockheed Martin Energy Research Corp. (Oak
Ridge National Laboratory), and  by the Polish Committee for
Scientific Research (KBN) under Contract No.~2~P03B~040~14.


\begin{thebibliography}{10}
\bibitem{sen} A similar conclusion can be drawn based on the
seniority model.
\end{thebibliography}
\end{document}